\documentclass[useAMS,usenatbib]{mnras}

\usepackage[T1]{fontenc}
\usepackage{pslatex}
\usepackage{amsmath}
\usepackage{amsfonts}
\usepackage{color}
\usepackage{url}
\usepackage{graphicx}
\usepackage{subfigure}
\usepackage{amssymb}
\usepackage{textcomp}
\usepackage{soul}
\usepackage{bm}
\usepackage{booktabs}
\usepackage{array}
\usepackage[draft]{hyperref}
 \graphicspath{{images/}}

{\newif\ifnotend
\notendtrue
\def\veclist{ABCDEFGHIJKLMNOPQRSTUVWXYZabcdefghijklmnopqrstuvwxyz.}
\def\top#1#2.{#1}
\def\tail#1#2.{#2.}
\loop\expandafter\xdef\csname v\expandafter\top\veclist\endcsname%
{{\noexpand\bf\expandafter\top\veclist}}
\edef\veclist{\expandafter\tail\veclist}
\if\veclist.\notendfalse\fi\ifnotend\repeat}

\mathchardef\mhyphen="2D

\title[Mass inflow rate into the Central Molecular Zone]{Mass inflow rate into the Central Molecular Zone: observational determination and evidence of episodic accretion}

\author[Sormani \& Barnes]{Mattia C. Sormani$^1$, Ashley T. Barnes$^{1,2}$ \\
$^1$Universit\"{a}t Heidelberg, Zentrum f\"{u}r Astronomie, Institut f\"{u}r theoretische Astrophysik, Albert-Ueberle-Str. 2, 69120 Heidelberg, Germany \\
$^{2}$Argelander Institute for Astronomy, University of Bonn, Auf dem H\''{u}gel 71, 53121 Bonn, Germany \\
}

\begin{document}

\date{}

\def\p{\partial}
\def\Omegap{\Omega_{\rm p}}

\newcommand{\co}{$\mathrm{^{12}CO}$}
\newcommand{\di}{\mathrm{d}}
\newcommand{\bfx}{\mathbf{x}}
\newcommand{\bfe}{\mathbf{e}}
\newcommand{\bfxi}{\bm{\xi}}
\newcommand{\vlos}{\mathrm{v}_{\rm los}}
\newcommand{\Tspin}{T_{\rm s}}
\newcommand{\Tb}{T_{\rm b}}
\newcommand{\degree}{\ensuremath{^\circ}}
\newcommand{\Th}{T_{\rm h}}
\newcommand{\Tc}{T_{\rm c}}
\newcommand{\bfr}{\mathbf{r}}
\newcommand{\bfv}{\mathbf{v}}
\newcommand{\bfu}{\mathbf{u}}
\newcommand{\pc}{\,{\rm pc}}
\newcommand{\kpc}{\,{\rm kpc}}
\newcommand{\Myr}{\,{\rm Myr}}
\newcommand{\Gyr}{\,{\rm Gyr}}
\newcommand{\kms}{\,{\rm km\, s^{-1}}}
\newcommand{\de}[2]{\frac{\partial #1}{\partial {#2}}}
\newcommand{\cs}{c_{\rm s}}
\newcommand{\rb}{r_{\rm b}}
\newcommand{\rqu}{r_{\rm q}}
\newcommand{\nuP}{\nu_{\rm P}}
\newcommand{\thetaobs}{\theta_{\rm obs}}
\newcommand{\hatn}{\hat{\textbf{n}}}
\newcommand{\hatt}{\hat{\textbf{t}}}
\newcommand{\hatx}{\hat{\textbf{x}}}
\newcommand{\haty}{\hat{\textbf{y}}}
\newcommand{\hatz}{\hat{\textbf{z}}}
\newcommand{\hatX}{\hat{\textbf{X}}}
\newcommand{\hatY}{\hat{\textbf{Y}}}
\newcommand{\hatZ}{\hat{\textbf{Z}}}
\newcommand{\hatN}{\hat{\textbf{N}}}
\newcommand{\hater}{\hat{\mathbf{e}}_r}
\newcommand{\hateR}{\hat{\mathbf{e}}_R}
\newcommand{\hatephi}{\hat{\mathbf{e}}_\phi}
\newcommand{\hatez}{\hat{\mathbf{e}}_z}
\newcommand{\hateP}{\hat{\mathbf{e}}_P}
\newcommand{\hatePhi}{\hat{\mathbf{e}}_\Phi}
\newcommand{\hatetheta}{\hat{\mathbf{e}}_\theta}
\newcommand{\hatemu}{\hat{\mathbf{e}}_\mu}
\newcommand{\hatenu}{\hat{\mathbf{e}}_\nu}
\newcommand{\hatePL}{\hat{\mathbf{e}}_{P\Lambda}}
\newcommand{\nablaPL}{\nabla_{P\Lambda}}

\def \Mdot {{\rm M$_\odot$}}
\def \Mdotyr {{\rm M$_\odot$\,yr$^{-1}$}}

\maketitle

\begin{abstract}
It is well known that the Galactic bar drives a gas inflow into the Central Molecular Zone, which fuels star formation, accretion onto the central super-massive black hole, and large-scale outflows. This inflow happens mostly through two symmetrical dust-lanes, similar to those often seen in external barred galaxies. Here we use the fact that the Milky Way dust-lanes have been previously identified in $^{12}$CO datacubes and a simple geometrical model to derive the first observational determination of the mass inflow rate into the Central Molecular Zone. We find that the time-averaged inflow rate along the near-side dust lane is $1.2^{+0.7}_{-0.8}\, \rm M_\odot yr^{-1}$ and along the far-side dust lane is $1.5^{+0.9}_{-1.0}\, \rm M_\odot yr^{-1}$, which gives a total inflow of $2.7^{+1.5}_{-1.7}\, \rm M_\odot yr^{-1}$. We also provide the time series of the inflow rate $\dot{M}$ for the future few Myr. The latter shows that the inflow rate is variable with time, supporting a scenario of episodic accretion onto the Central Molecular Zone.
\end{abstract}

\begin{keywords}
Galaxy: nucleus - Galaxy: centre - ISM: kinematics and dynamics - Galaxy: kinematics and dynamics - galaxies: star formation
\end{keywords}

\section{Introduction}

Determining the mass inflow rate driven by the Galactic bar into the Central Molecular Zone (CMZ) is important for a number of reasons. This inflow is what created the mass concentration known as the CMZ in the first place, and it determines its star formation rate \citep[e.g.][]{Longmore+2013,Kruijssen+14b}. It affects the chemical and dynamical evolution of the stellar bulge/bar over secular time-scales \citep[e.g.][]{Norman+1996,KormendyKennicut2004,Cole+2014,Debattista+2017}. The inflow may fuel the super-massive black hole at the centre, although it is currently unclear how the gas migrates from the CMZ ($R\sim10^2\pc$) down to the accretion disc at much smaller radii ($R\sim10^{-3}\pc$) \citep[e.g.][]{Phinney1994,HopkinsQuataert2010,Emsellem+2015,Li+2017}. Finally, the gas acts as a fuel for the outflow associated with the Fermi Bubbles \citep{BHCohen2003,Su+2010}.

Theoretical studies and observations of external galaxies indicate that the accretion happens mostly through two symmetrical dust-lanes, along which the gas streams from distances of $R\sim 3$ kpc directly down towards the CMZ \citep[e.g.][]{Athan92b,Athanassoula1994,Regan+1997,Elmegreen+2009,Kim++2012a,SBM2015a,Sormani+2018}. The dust-lanes of the Milky Way (MW) have been identified as two prominent features in the CO $(l,b,v_{\rm los})$ datacubes \citep[e.g.][]{Fux1999, Marshall2008, Li++2016, Sormani+2018} and can be seen in extinction maps derived from infrared observations \citep{Marshall+2006,Marshall2008}. Hence one might expect that by coupling theoretical understanding with the features identified in the data, it should be possible to obtain an estimate of the inflow rate.

In this paper we use a simple geometrical model, inspired by theoretical studies of gas flow in barred potentials, to obtain the first observational determination of the inflow rate into the CMZ.

\section{Methods}

\subsection{Identification of dust lanes}

\begin{figure}
\centering
\includegraphics[trim = 16mm 10mm 9mm 2mm, clip, width=1\columnwidth]{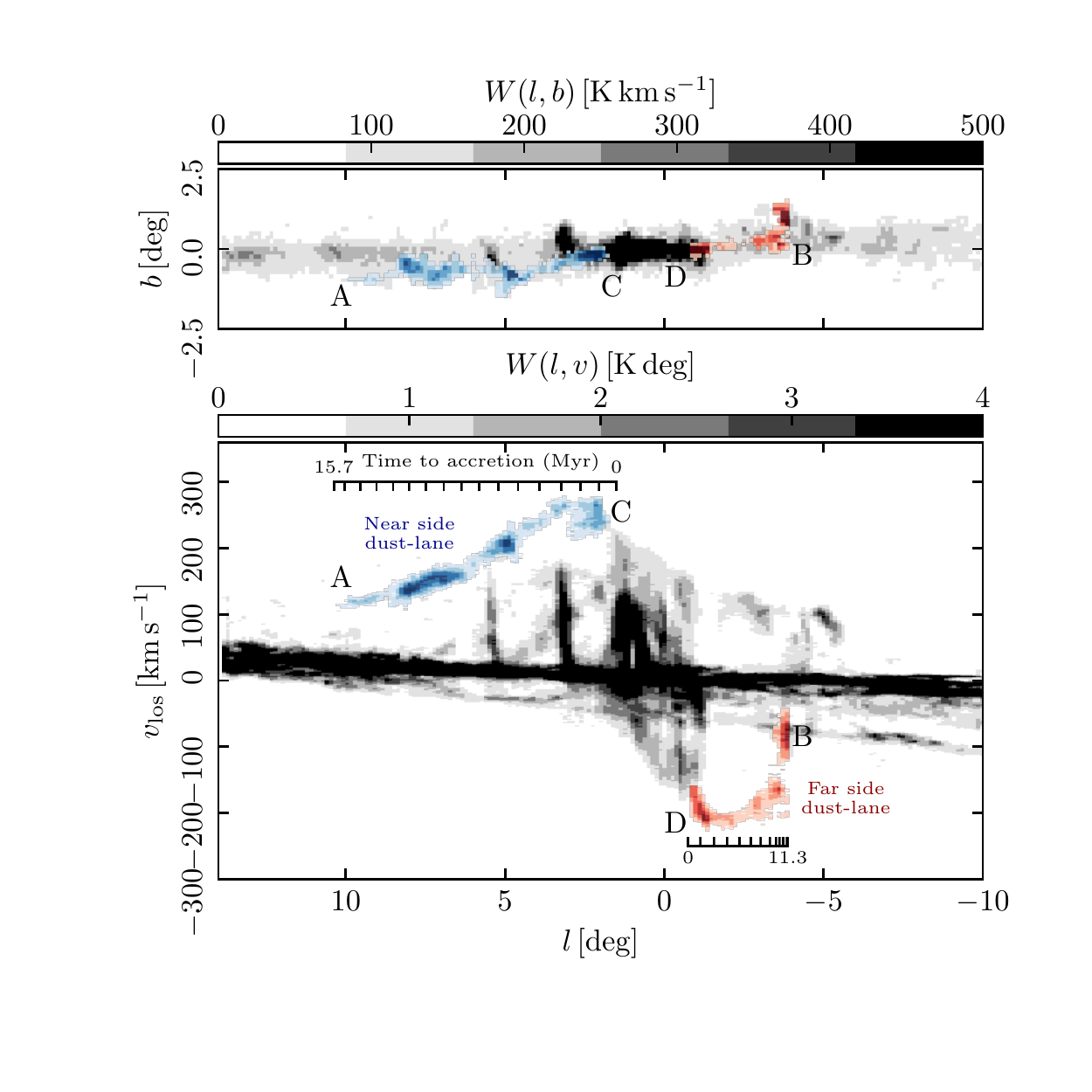}
\caption{The \co\ emission associated with the Galactic bar dust lanes, where the near-side and far-side are shown in blue and red, respectively. These are have identified with the procedure outlined by \citet{Marshall2008}, using the data taken from \citet{COdata}. The upper panel shows the \co\ emission integrated in velocity, which is used to determine the mass, and the bottom panel shows the emission integrated in latitude. Note that the colour scales of the dust lanes (red and blue) and of the background emission (grey) are different and have been adjusted to improve visualisation. The alphabetical labels correspond to the points shown in Figure \ref{fig:2}.}
\label{fig:1}
\end{figure}

Two features have been identified in $(l,b,v_{\rm los})$ datacubes as the dust lanes of the MW bar. The first is, for historical reasons, often called the `connecting arm' \citep[e.g.][]{CohenDavies1976,RodriguezFernandez+2006} and is shown in blue in Fig. \ref{fig:1}. It corresponds to the near-side dust lane and is visible in both HI and CO large-scale surveys. The second corresponds to the far-side dust lane and is shown in red in Fig. \ref{fig:1}. The interpretation of these features as the dust lanes associated with the MW bar was first put forward by \cite{Fux1999}\footnote{It is worth mentioning for historical completeness that in the very early days \cite{Kerr1967} speculated about the existence of a bar (at the time it was not known whether the MW had a bar) and using HI data interpreted the connecting arm as an associated dust lane, although contrary to the current interpretation he placed it on the far side.} (see also \citealt{Li++2016,Sormani+2018}) by comparing synthetic $(l,b,v_{\rm los})$ datacubes from gas dynamical simulations to observations, and independently confirmed by \cite{Marshall+2006,Marshall2008}, who identified the dust lanes in three dimensional extinction maps of the inner Galaxy created using the Two Micron All Sky Survey (2MASS) \citep{Skrutskie+2006}.

In this paper we define the dust-lane features  by following the same procedure described in \citet{Marshall2008}. We use the $J = 1\,\rightarrow\,0$ \co\ data of \citet{COdata}, which contains a multitude of complex emission features along each the line-of-sight. \citet{Marshall2008} imposed three selection criteria in $(l,b,v_{\rm los})$ space in order to isolate the dust-lane emission. In particular: (i) the emission is restricted to $(l,v_{\rm los})$ regions that enclose the features (see upper panel in their fig. 3), while being careful in excluding emission coming from the nuclear ring/disc \citep{Molinari+2011} as well as other known sources (e.g. Bania's Clump 1 and 2, \citealt{Bania1977,StarkBania1986}); (ii) the emission is restricted in $(l,b)$ to exclude some emission that appears to be distinct and not associated with the near-side dust lane (see lower-left panel in fig. 3 of \citealt{Marshall2008}); (iii) the emission is further restricted in $(b,v_{\rm los})$ space to avoid contamination from a prominent feature known as the 3kpc arm \citep[e.g.][]{COdata} and other foreground emission (see lower-right panel in fig. 3 of \citealt{Marshall2008}). Figure \ref{fig:1} shows the result of this procedure.

\subsection{Determination of masses}

We assume that the $J = 1\,\rightarrow\,0$ transition of \co \, is a diagnostic tracer of the bulk molecular gas, i.e. we assume it is proportional to the total gas mass. The integrated intensity of CO, $W(l, b)$, is first converted to H$_2$ column density, $N(\mathrm{H_2})$, using an $X_{\rm CO}$ conversion factor of $X_{\rm CO} = 2 \times 10 ^{20} (N(\mathrm{H_2}) \, \mathrm{cm}^{-2}) \, / \, \mathrm{(}W(l, b) \, \mathrm{ K~km s^{-1} )}$ (e.g. \citealp{bolatto_2013}). We discuss the uncertainty associated with this conversion factor below in Sect. \ref{sec:XCO}. Then we convert the H$_2$ column density to total mass by using a mean mass per particle of 2.8 atomic mass units (which takes into account the mass contribution from helium and metals, see appendix of \citealp{kauffmann_2008}) and the distance derived using the geometry described in the following section. This procedure gives the total mass associated with the dust lanes in each pixel in $(l,b,v)$ space.

As a sanity check, we have also estimated the mass in neutral $\rm HI$ contained in the dust lane features using data from the LAB survey \citep{HIdata}. For the near-side dust lane we found a mass of approximately $\simeq 1\times 10^{6} \, M_\odot$, in good agreement with that found for the same feature by \cite{CohenDavies1976}. This mass gives a negligible contribution compared to the errors in the molecular mass resulting from the uncertainty in the $X_{\rm CO}$ factor, so we have ignored it in our calculations.

\subsection{Geometry of the model}

\begin{figure}
\centering
\includegraphics[width=0.4\textwidth]{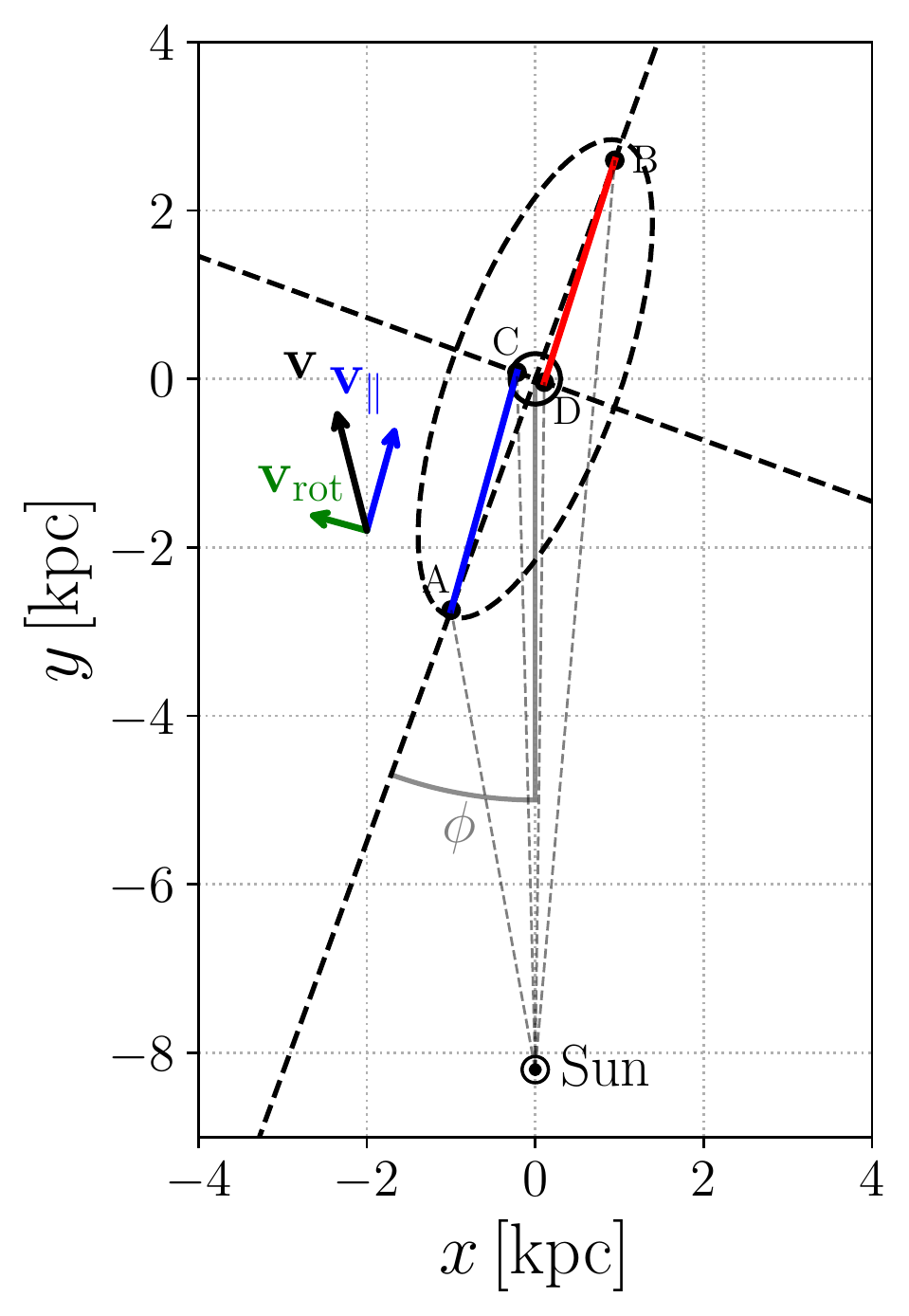}
\caption{The geometry of our model. The blue and red segments represent the dust lanes. The dashed ellipse schematically represents the Galactic bar, and the dashed black lines its minor and major axis. The points A,B (C,D) are the interception of the maximum (minimum) absolute longitudes at which the dust lanes are visible in the data with the major (minor) axis of the bar. The inner circle of radius $R_{\rm CMZ}=300\pc$ schematically represents the CMZ.}
\label{fig:2}
\end{figure}

To derive the mass inflow rate $\dot{M}(t)$ we need the time $t(l)$ that it takes for a point on the dust lane at longitude $l$ to reach the CMZ. In order to do this, we need to deproject the dust lanes from the observational space, $(l,b,v_{\rm los})$, to Galactocentric coordinates $(x,y,v_x,v_y)$. This is a degenerate problem, but we can rely on what we know about the geometry of the dust lanes and the gas dynamics in barred potentials from theoretical studies and observations of external barred galaxies. We use the following procedure (see Fig. \ref{fig:2}):
\begin{itemize}
\item We assume that the angle between the major axis of the bar and the Sun-Galactic centre line is $\phi=20\degree$, consistent with various independent estimates from gas dynamical modelling, star counts, and near infrared photometry \citep{Fux1999,BissantzGerhard2002,Cao+2013,Wegggerhard2013}. We later explore the effects of varying the angle in the range $\phi=15\mhyphen30\degree$. We assume that the Galactocentric position of the Sun is $ \bfr_\odot = (0, -8.2 \kpc)$ and its velocity is $\bfv_\odot = (-240\kms,0)$ \citep[e.g.][]{BlandhawthornGerhard2016}.
\item We assume that the dust lanes are two straight segments. The dust lanes are assumed to intersect the major axis of the bar at their furthest from the Galactic centre (point A and B in Fig. \ref{fig:2}) and the minor axis of the bar at their extreme closest to the Galactic centre (point C and D in Fig. \ref{fig:2}). 
\item The dust lanes are assumed to be stationary in the frame rotating with the bar. The bar is assumed to rotate rigidly with a constant pattern speed of $\Omega_{\rm p}=40\kms \kpc^{-1}$, as implied by recent estimates \citep{SBM2015c,Portail+2017,Perez+2017}.
\item Theoretical studies show that, in the frame of the bar, the velocity of the gas is approximately parallel to the dust lanes, and that the gas plunges along the dust lanes directly from Galactocentric distances of $R\sim 3\kpc$ down to the CMZ \citep[e.g.][]{Athan92b,SBM2015a,Sormani+2018}. Under this assumption, the velocity of the gas in the frame rotating with the bar can be written as $\bfv_\parallel= v_\parallel \hat{\mathbf{e}}_\parallel$, where $\hat{\mathbf{e}}_\parallel$ is the unit vector parallel to the dust lane, and can be related to the observed line-of-sight velocity $v_{\rm los}$ in the following way. The total velocity of a gas parcel falling down the dust lane in the inertial frame of the Galaxy is
\begin{equation}
\bfv = \bfv_\parallel + \bfv_{\rm rot},
\end{equation}
where $\bfv_{\rm rot} = \Omega_{\rm p} \hatez \times \bfr $ is the rotational velocity due to the fact that the bar rigidly rotates with a constant pattern speed, $\hatez$ is the unit vector perpendicular to the plane of the Galaxy and $\bfr$ is the Galactocentric radius. The unit vector in the direction of the line-of-sight is $\hat{\mathbf{e}}_{\rm los}= ({ \bfr - \bfr_\odot }) / {|  \bfr - \bfr_\odot |}$. The line of sight velocity is then $v_{\rm los} = (\bfv - \bfv_\odot) \cdot \hat{\mathbf{e}}_{\rm los}$. Isolating $v_\parallel$ yields
\begin{equation} \label{eq:vpar}
v_\parallel = \frac{v_{\rm los} - (\bfv_{\rm rot} - \bfv_\odot)\cdot \hat{\mathbf{e}}_{\rm los}}{\hat{\mathbf{e}}_\parallel \cdot \hat{\mathbf{e}}_{\rm los}}.
\end{equation}
This relation allows to obtain $v_\parallel$ given the observed $v_{\rm los}$. For each longitude $l$,  the dust lanes have a spread in $v_{\rm los}$ (see Fig. \ref{fig:1}), hence we use in Eq. \ref{eq:vpar} the mass-weighted value of $v_{\rm los}$ at that value of longitude.
\end{itemize}
This completes the specification of our geometrical model. From the first two items above we can obtain the position and distance of each point along the dust lanes, which is used to determine the gas mass (see previous section). Then from $v_{\parallel}(l)$ we can derive the time $t(l)$ that it takes for a parcel of gas to reach the end of the dust lane.

\section{Results} \label{sec:results}

\begin{figure}
\centering
\includegraphics[width=0.5\textwidth]{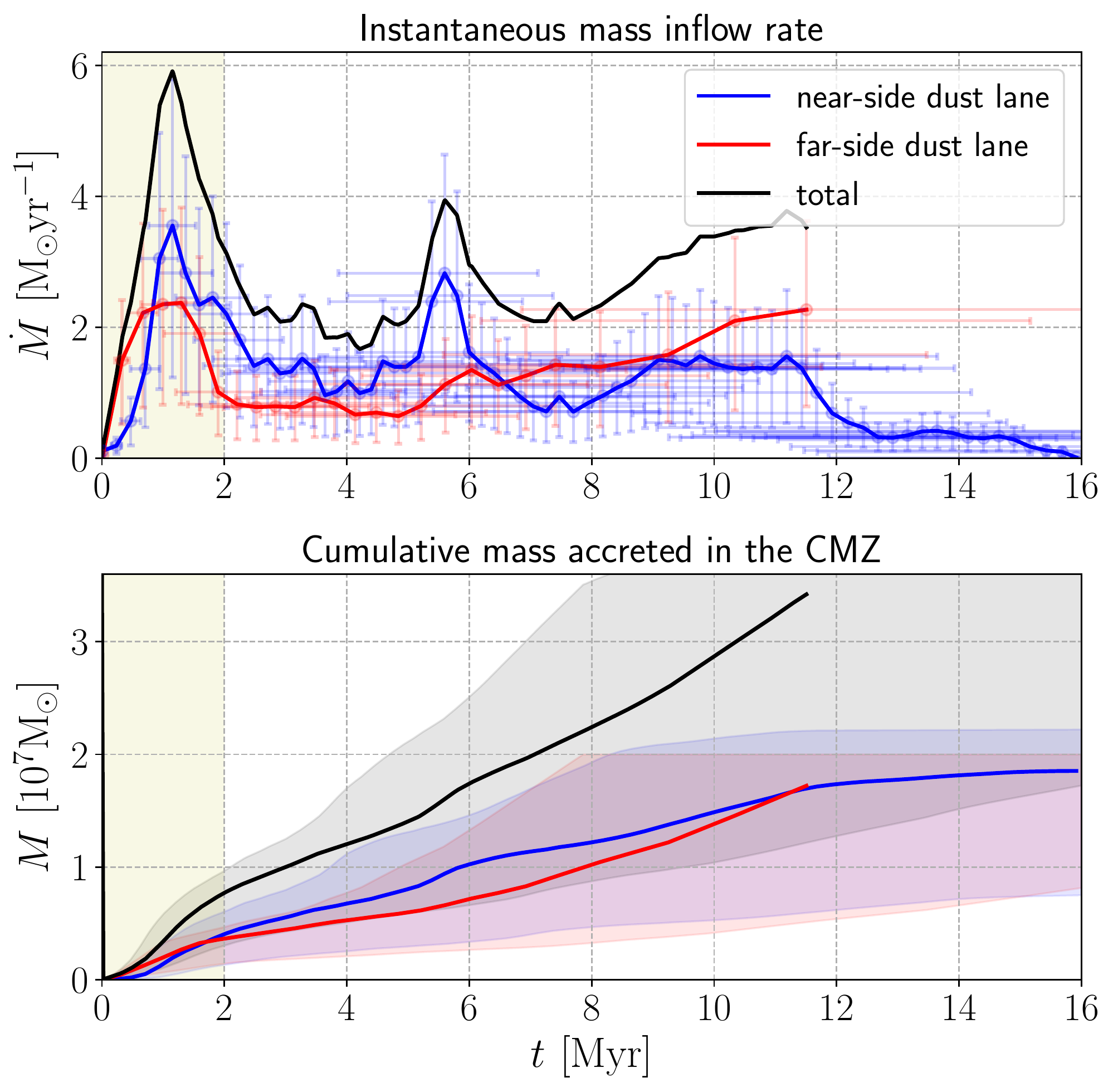}
\caption{The instantaneous (upper panel) and cumulative (lower panel) inflow rate into the CMZ calculated for the future several Myr. The near- and far-side dust lanes are displayed as blue and red lines, respectively. The error bars in the upper panel and the shaded regions in the lower panel show the total uncertainty associated with (i) varying the angle between the major axis of the bar and the Sun-Galactic centre line in the plausible range $\phi = 15 \mhyphen 30 \degree$ around the fiducial value of $\phi=20\degree$ and (ii) the $X_{\rm CO}$ factor. The yellow shaded regions on the left shows where our results could be affected by interaction with the CMZ (see Section \,\ref{sec:results}).}
\label{fig:3}
\end{figure}

The upper panel of Figure \ref{fig:3} displays the instantaneous mass inflow rate into the CMZ as a function of future time for the material currently on the two dust-lanes, while the lower panel displays the cumulative accreted mass. We find that the time-averaged accretion rate from the near-side dust lane is $1.2^{+0.7}_{-0.8}\, \rm M_\odot yr^{-1}$ and along the far-side dust lane is $1.5^{+0.9}_{-1.0}\, \rm M_\odot yr^{-1}$, which gives a total inflow of $2.7^{+1.5}_{-1.7}\, \rm M_\odot yr^{-1}$. The errors quoted here and shown in Fig.~\ref{fig:3} reflect two sources of uncertainties. The first is the uncertainty in the geometry associated with the angle $\phi$ between the major axis of the bar and the Sun-Galactocentric line, which is varied in the plausible range $\phi = 15\mhyphen30 \degree$ \citep[e.g.][]{BlandhawthornGerhard2016}. The second is the uncertainty in the $X_{\rm CO}$ factor. The determination of the uncertainties is discussed in more detail in Sect. \ref{sec:errors}.

An interesting point regarding the upper panel of Figure \ref{fig:3} is the large variability that is observed around the time-averaged inflow rates of $1.2\, \rm M_\odot yr^{-1}$ and $1.5\, \rm M_\odot yr^{-1}$. The blue and red curve show several peaks associated with stronger inflow, reaching twice or thrice the average inflow, with quieter periods in between. We note that for $t \gtrsim 2 \Myr$ these peaks are clearly associated with brighter emission along the dust lanes (see for example the blue emission in Fig \ref{fig:1} which peaks around $l\sim4.5\degree,v_{\rm los}\sim150\kms$ and $l \sim 7 \degree,v_{\rm los}\sim200\kms$), hence it is likely that they are real and not an artefact of our model. For $t < 2 \Myr$ (yellow shaded areas in Fig. \ref{fig:3}), however, both the near- and far-side dust lanes show a peaked emission, which is a suspicious coincidence. Closer inspection of Fig. \ref{fig:1} shows that some of the emission corresponding to those regions is connected with emission coming from the CMZ. Hence it is possible that our results for the first 2 Myr are biased by the fact that the gas on the dust lanes has started interacting with the CMZ. Since the gas that has been falling with a high velocity down the dust lanes strongly shocks as it crashes into slower material in the CMZ, it will tend to have a very large local velocity dispersion, as can already be seen in Figure \ref{fig:1}. If the interaction does not significantly affect the molecular content of the gas, this will lead to a decrease in $X_{\rm CO}$ compared to the value we assume here \citep[see e.g.][]{Shetty+2011}, leading to an overestimate of the gas mass and therefore spurious peaks.

The variations in the inflow rate are interesting as they clearly highlight the clumpy nature of the inflow into the CMZ. This could have important implications for the many processes that are driven by this inflow (see introduction and Sects. \ref{sec:wherego} and \ref{sec:turbulence}), the most direct of which is probably the star formation rate. Previous studies have shown that the star formation rate within this region has been constant within a factor of a few for the past few \Myr\ (e.g. \citealp{barnes_2017}), which is consistent with the periodicity and duration of the large peaks in the instantaneous mass inflow rate, see for example the peak at $\sim$\,5$\Myr$ and $\sim$\,10$\Myr$ in the blue line in the upper panel of Fig. \ref{fig:3}. Indeed, it has been suggested that the CMZ region undergoes a cycle containing periods of quiescence and rapid star formation (albeit invoking a different mechanism and on a slightly longer timescale than is observed here; \citealp{Kruijssen+14b, KrumholzKruijssen2015, KrumholzKruijssen2017}).

\section{Discussion}\label{sec:caveats}

\subsection{Sources of errors} \label{sec:errors}

The errors quoted above reflect two sources of uncertainties: (i) the bar angle $\phi$ and (ii) the $X_{\rm CO}$ factor. In this section we discuss these and some additional sources of error that could potentially affect the results presented here.

\subsubsection{Bar angle $\phi$}

To obtain the error associated to the geometry of our model, we have varied the angle $\phi$ between the major axis of the bar and the Sun-Galactocentric line in the plausible range $\phi = 15\mhyphen30 \degree$ \citep[e.g.][]{BlandhawthornGerhard2016}. If a smaller (larger) angle is assumed, our geometrical model yields longer (shorter) dust-lanes, which increases (decreases) the time it takes for the same parcel of gas to reach the centre and decreases (increases) the inflow rate $\dot{M}$. Thus, if $\phi$ is decreased (increased) the inflow rate also decreases (increases), and the plots in Fig. \ref{fig:3} are stretched (compressed) in the horizontal direction while approximately preserving the area beneath them. Note that changing $\phi$ also affects the time for a gas parcel to reach the CMZ, and this is the origin of the horizontal errors bars in Fig. \ref{fig:3}. 

Accounting only for the uncertainty on $\phi$, we would get $1.2^{+0.6}_{-0.3}\, \rm M_\odot yr^{-1}$, $1.5^{+0.8}_{-0.4}\, \rm M_\odot yr^{-1}$, $2.7^{+1.3}_{-0.7}\, \rm M_\odot yr^{-1}$. Note that since the errors on the two dust lanes are correlated, the error on the total accretion rate is the sum of the two rather than the sum in quadrature (which would assume the errors are independent).

\subsubsection{$X_{\rm CO}$} \label{sec:XCO}

The value of the $X_{\rm CO}$ factor chosen here is an average taken from various measurement methods (e.g. virial equilibrium method, optically thin tracers, dust extinction, dust emission, gamma-rays)  for the molecular clouds within the disc of the MW (see \citealp{bolatto_2013}). This value is found to vary significantly on a cloud-to-cloud basis, reflecting local chemistry and physical conditions, and thus carries an uncertainty that \cite{bolatto_2013} estimated to be $\pm\,30\%$. However, the following considerations suggest that in the case of the specific features studied in this paper the uncertainty should be even higher. 

First, in a highly dynamic environment such as that present in the dust lanes, the assumption of virial equilibrium which forms the basis of the virial method is probably not appropriate: a typical molecular cloud would not have time to reach equilibrium before being sheared out. The fact that the other methods mentioned above, which are independent of the virial assumption, all point towards approximately the same value for the $X_{\rm CO}$ factor is somewhat reassuring, but this should be tested specifically on the dust lanes features. 

Second, the $X_{\rm CO}$ averaged over the MW disc may not be appropriate for the dust lanes if the $X_{\rm CO}$ varies systematically with Galactocentric radius. There is evidence that the $X_{\rm CO}$ factor is lower by factors of several in the very central region of the MW (i.e. at Galactocentric distances of $<$\,500\,pc, e.g. \citealp{sodroski_1995, oka_1998, oka_2001, strong_2004, ackermann_2012}). However, the evidence for a large-scale Galactic $X_{\rm CO}$ gradient between the Solar neighbourhood ($R=8\kpc$) and the outer tip of the dust lanes ($R\sim3\kpc$) is currently inconclusive \citep[e.g.][]{bolatto_2013}. Given that a complete determination of the uncertainties discussed here is beyond the scope of this paper, we apply a constant conversion factor and anticipate a positive uncertainty comparable to the variation observed for disc star-forming regions (i.e. +30\%), and a negative uncertainty a factor of two larger (i.e. -60\%) to account for the possible decrease of the $X_{\rm CO}$ factor with Galactocentric radius.

Accounting only for the $X_{\rm CO}$ uncertainty, we would get $1.2^{+0.4}_{-0.7}\, \rm M_\odot yr^{-1}$, $1.5^{+0.4}_{-0.9}\, \rm M_\odot yr^{-1}$, $2.7^{+0.8}_{-1.6}\, \rm M_\odot yr^{-1}$. These errors have been added in quadrature to the errors associated with $\phi$ to obtain the total uncertainty.

\subsubsection{Definition of dust lane features}

The dust lanes features are quite isolated in the $(l,v)$ plane (Fig.~\ref{fig:1}) and thanks to their high $v_{\rm los}$ there is little overlapping material that could be confused with them (foreground emission belonging to the Galactic plane mostly lies at much smaller velocities). However, due to the limited resolution of the \cite{COdata} data it is possible that our selection in the \co\, datacube contains some spurious gas not belonging to the dust lanes, or that we are missing some gas that it actually does belong to them. Higher-resolution data could help to better isolate the relevant features \citep[e.g.][]{Schuller+2017} and improve our estimation of their masses.

\subsubsection{Overshooting} \label{sec:overshooting}

Another possible source of error is the phenomenon of `overshooting'. Simulations show that sometimes the gas falling fast down the dust lanes, instead of crashing into the CMZ, misses it and `overshoots', eventually hitting the dust lane on the opposite side \citep[e.g.][]{Sormani+2018}. This effect could be taken into account in a simple way including a factor $f \leq 1 $ which quantifies the fraction of gas that is deposited in the CMZ. This is a sort of `effective cross section' and would change our results by a scaling factor. One could estimate $f$ by performing hydrodynamic simulations and using tracer particles to follow molecular clouds while they are falling along the dust lanes to find what fraction of their mass is deposited into the CMZ. However, particular care must be taken in modelling the equation of state of the interstellar medium correctly \citep[see the discussion in section 4 of][]{Elmegreen+2009} and in dealing with the transients associated with the `gradual turn on' of the bar that is typically used in these simulations \citep[e.g.][]{Athan92b,Sormani+2018}. These investigations are beyond the scope of the current paper. Here, rather than include an unjustified value of $f$, we prefer not to apply any correction. In this sense, our estimates for the inflow can be considered as upper limits.

\subsection{Where does the gas go?} \label{sec:wherego}

The total molecular gas mass in the CMZ is of the order of $5\times~10^7~ \rm M_\odot$ \citep{Dahmen+1998,PiercePrice+2000}. At an inflow rate of $\dot{M} =  2.7\, \rm M_\odot yr^{-1}$, this would take only $\sim 20 \Myr$ to build up. This is much smaller than the age of the Galactic bar \citep[e.g.][]{Debattista+2018,Buck+2018}, and if our determined value of the current inflow rate is in any way representative, something must be getting rid of most of the gas which falls in.

The current star formation rate (SFR) of the CMZ is estimated to be of the order of $\sim 0.1 \, \rm M_\odot yr^{-1}$ \citep[e.g.][]{YusefZadeh2009,Immer+2012b,Longmore+2013}, and there is evidence that this rate has been constant during the past few Myr \citep{barnes_2017}. Therefore, these values suggest that only a small fraction ($\sim 5\%$) of the infalling gas is turned into stars.

Observations show that a substantial amount of gas leaves the nuclear regions through the outflow associated with the Fermi Bubbles, but the total amount of outflowing gas is very uncertain. By modelling the kinematics of 21cm $\rm HI$ datacubes, \cite{Diteodoro+2018} estimated an outflow in {\it neutral HI gas} of $\dot{M}_{\rm HI} \sim 0.1 \rm M_\odot yr^{-1}$. By modelling the kinematics of UV absorption spectra of multiple background sources, \cite{Bordoloi+2017} found an outflow in {\it warm ionised gas} of $\dot{M}_{\rm WIM} \gtrsim 0.4 \rm M_\odot yr^{-1}$.\footnote{The estimate given in their paper of $\dot{M} \gtrsim 0.2 \rm M_\odot yr^{-1}$ only concerns the Northern Fermi bubble. Hence, assuming reflection symmetry about the Galactic plane, we have multiplied by a factor of 2 to obtain the total outflow.} To the best of our knowledge the outflow in {\it cold molecular gas} is currently unknown, but we might expect that it is at least comparable to that in neutral HI gas, $\dot{M}_{\rm cold} \gtrsim 0.1 \rm M_\odot yr^{-1}$. Using X-ray {\rm OVII} and {\rm OVIII} line observations, \cite{MillerBregman2016} estimated the hot gas mass within the Fermi bubbles to be $M_{\rm hot} \simeq 10^7 \, \rm M_\odot$ and an expansion velocity of $v_{\rm exp} \sim 500 \kms$, although these numbers have large uncertanties. A further uncertainty comes from the fact that we do not know what fraction of the gas in $M_{\rm hot}$ actually comes from the nuclear region of the Galaxy and what fraction is shocked gas belonging to the hot gaseous corona (i.e., the circum-galactic medium) which has mixed with outflowing gas. If the hot gas within the bubbles all comes from the nuclear regions, using the above values and assuming that gas composing the Fermi Bubbles travels an average distance of $d\sim 5\kpc$ from the Galactic plane one gets an outflow rate of $\dot{M}_{\rm hot} \simeq M_{\rm hot} v_{\rm exp} / d \sim 1.0 \, \rm M_\odot yr^{-1} $, which should be considered as an upper bound. Putting all these measurements together one gets a total outflow estimate in the range $\dot{M} \sim 0.7 \mhyphen 1.7 \, \rm M_\odot yr^{-1}$. These values are consistent with our inflow estimates within the errors. 

To summarise, it appears that most of the gas inflowing into the CMZ is eventually expelled through the outflow associated with the Fermi bubbles, while a fraction that can be as small as $\sim 5\%$ is turned into stars. However, if star formation is episodic and the current value for the SFR is a near-minimum of a longer ($10\mhyphen 20\Myr$) star-formation cycle \citep[e.g.][]{Kruijssen+14b}, then star formation may give a much larger contribution on average. The total observed gas consumed in star formation + Fermi bubbles outflow is of the order of $\lesssim 1 \, \rm M_\odot yr^{-1}$, which although consistent with our values within the errors it could be an indication that, as a consequence of neglecting the `overshooting' effect (see Sect. \ref{sec:overshooting}), our estimate for the inflow rate may well be an overestimate.

\subsection{Where does the gas come from?} \label{sec:wherecome}

It is well known that the observed depletion in the radial distribution of molecular gas in the central $R \lesssim 4\kpc$ of our Galaxy \citep[e.g.][]{HeyerDame2015} is caused by the bar which clears the area and causes the gas to flow inwards. How much time does the bar need to clear the $R \lesssim 4 \kpc$ region? If we imagine extrapolating the surface density in fig. 7 of \cite{HeyerDame2015} into the centre, the original mass contained in the region would be $M(R<{4\kpc}) \simeq \pi (4 \kpc)^2 \times 5\, \rm M_\odot \, pc^{-2} \simeq 2.5 \times 10^8 \, \rm M_\odot$. At an inflow rate of $\dot{M} =  2.7\, \rm M_\odot yr^{-1}$, this would take only $\sim 100  \Myr$ to clear. There are two possibilities: 

\begin{enumerate}
\item Something is replenishing the reservoir of gas that supplies the bar inflow. The most likely possibility is that the gas is transported radially within the disc \citep[e.g.][]{LaceyFall1985,BilitewskiSchoenrich2012,Cavichia+2014,Kubryk+2015a,Kubryk+2015b}. Proposed mechanisms include (a) raining of gas with low angular momentum from the circumgalactic medium which, by mixing with gas in the disc, causes the latter to move inwards (b) viscous accretion (c) interaction of the gas with bar/spiral patterns. However, the relative contribution of these three items is currently poorly understood. There are very little direct observational constraints on the amount of gas raining from the circumgalactic medium, and viscous torques seem to be negligible. In relation to mechanism (c), one needs to explain how the gas crosses the corotation `barrier' (which for the MW is at approximately at $R \simeq 6 \kpc$ assuming a bar pattern speed of $\simeq 40 \kms \kpc^{-1}$, see for example table 3 of \citealt{SBM2015c}) which is believed to prevent gas from outside corotation to reach inside corotation. It has been proposed that interaction between the bar and a spiral pattern with a different/no pattern speed may help the gas overcome the barrier \citep[][]{Gerhard2011}. The fact that the corotation radius increases over time (as a consequence of the pattern speed of the bar decreasing over time due to secular evolution, e.g. \citealt{Athanassoula2003,Wu+2018}), thereby increasing the amount of gas available for accretion, may also play a role \citep[on this point see also][]{Elmegreen+2009}.
\item  The gas reservoir that supplies the bar inflow is not replenished. In this case the bar must be extremely young. Even if we are overestimating the inflow rate by a factor of $\sim10 $ (a factor larger than this would not be consistent with the observed SFR and/or Fermi bubbles outflow, see Sect. \ref{sec:wherego}), the clear up time would be of the order of $\sim 1 \Gyr$, still significantly smaller than the commonly accepted value for the age of the bar \citep[e.g.][]{Debattista+2018,Buck+2018}.
\end{enumerate}
Therefore, unless the bar is much younger than currently believed ($t_{\rm bar} \lesssim 1 \mhyphen 2 \Gyr$), some gas replenishment must take place in the region just outside the bar. This may be considered an indirect evidence for the presence of radial flows within the disc of our Galaxy which bring the gas down to the outskirts of the Galactic bar ($R\sim4\kpc$). The mechanism by which this happens is unclear and deserves further investigation.
 
\subsection{Impact on driving CMZ turbulence} \label{sec:turbulence}

One of the open questions regarding the CMZ is what drives the turbulence \citep[e.g.][]{Kruijssen+14b}. Using our derived value for the mass inflow, we can make a simple estimate of its impact in driving the turbulence. The kinetic energy per unit time deposited into the CMZ is of order $\dot{M} v_{\rm inflow}^2/2\sim 8 \times 10^{39} \, \rm erg \, s^{-1}$, where we have used our fiducial value $\dot{M} =  2.7\, \rm M_\odot yr^{-1}$ and we have taken $v_{\rm inflow} \sim 100 \kms$ as a representative value for the relative velocity between the infalling gas and the gas already in the CMZ. The energy per unit time dissipated by the observed turbulent motions can be estimated as $\dot{E} \sim M_{\rm CMZ} \sigma^3 / h \sim 5 \times 10^{39} \, \rm erg \, s^{-1} $ \citep[e.g.][]{MaclowKlessen2004}, where we have used a total CMZ mass of $M_{\rm CMZ} = 5 \times~10^7\, \rm M_\odot$, an observed velocity dispersion of $\sigma = 20 \kms$ and a CMZ scale-height of $h \sim 50 \pc$. Thus we find that gas inflow is a promising candidate for driving the turbulence in the CMZ. Compared to \cite{Kruijssen+14b}, who made a similar estimate but in the absence of an available measurement for the inflow rate used a lower value than we found in this paper, our results suggest that the importance of inflow-driven turbulence to the overall energy budget is more important than previously assumed.

\section{Conclusion}

We have presented the first observational determination of the accretion rate into the Central Molecular Zone. By using a simple geometrical model, we have determined the time-averaged inflow to be $1.2^{+0.7}_{-0.8}\, \rm M_\odot yr^{-1}$ and $1.5^{+0.9}_{-1.0}\, \rm M_\odot yr^{-1}$ along the near- and far-side dust lanes respectively, giving a total inflow of $2.7^{+1.5}_{-1.7}\, \rm M_\odot yr^{-1}$. The main uncertainty lies in the $X_{\rm CO}$ conversion factor. Other findings can be summarised as follows:
\begin{enumerate}
\item We have found evidence for time-variability and clumpy nature of this inflow, which suggests that accretion is episodic, with potentially interesting consequences for the star formation cycle in the CMZ.
\item  It appears that most of the gas inflowing into the CMZ is eventually expelled through the outflow associated with Fermi bubbles, while a fraction that can be as small as $\sim 5\%$ is turned into stars. However, if star formation is episodic and the current value for the SFR is a near-minimum of a longer ($10\mhyphen 20\Myr$) star-formation cycle \citep[e.g.][]{Kruijssen+14b}, then star formation may give a much larger contribution on average. 
\item Unless the bar is much younger than commonly accepted ($t_{\rm bar} \lesssim 1 \mhyphen 2 \Gyr$), something must be replenishing the reservoir of gas that supplies the bar inflow. This may be considered an indirect evidence for the presence of radial flows within the Galactic disc. However, the precise mechanism by which this happens is unclear.
\item The kinetic energy provided by the inflowing gas seems sufficient to drive the observed turbulence in the CMZ.
\end{enumerate}

\section*{Acknowledgements}

The authors are grateful to Bruce Elmegreen for extensive and insightful comments and discussions. We thank Cara Battersby, James Binney, Enrico Di Teodoro, Adam Ginsburg, Simon Glover, Perry Hatchfield, Jonathan Henshaw, Steven Longmore, Diederik Kruijssen, John Magorrian, Ralph Schoenrich and Emanuele Sobacchi for useful comments on an earlier draft of this paper. We are grateful to the anonymous referee for useful and constructive comments that improved the quality of the paper. MCS acknowledges support from the Deutsche Forschungsgemeinschaft via the Collaborative Research Centre (SFB 881) ``The Milky Way System'' (sub-projects B1, B2, and B8). ATB would like to acknowledge the funding provided by the European Union's Horizon 2020 research and innovation programme (grant agreement No 726384). This research made use of the Glue software package \citep{beaumont_2015, robitaille2017}, and the Astropy package, a community-developed core Python package for Astronomy (Astropy Collaboration, 2018). 

\def\aap{A\&A}\def\aj{AJ}\def\apj{ApJ}\def\mnras{MNRAS}\def\araa{ARA\&A}\def\aapr{Astronomy \&
 Astrophysics Review}\def\apjs{ApJS}\def\apjl{ApJ}\def\pasj{PASJ}\def\nat{Nature}\def\prd{Phys. Rev. D}
\def\ssr{Space Sci. Rev.}\def\pasp{PASP}\def\pasa{Publications of the Astronomical Society of Australia}
\bibliographystyle{mnras}
\bibliography{bibliography}

\end{document}